\documentclass[a4paper, 12pt]{article}

\usepackage{a4wide}
\usepackage[utf8]{inputenc}
\usepackage[english]{babel} 
\usepackage{amsmath}
\usepackage{amsfonts}
\usepackage{amssymb}
\usepackage{amsthm}
\usepackage{amscd}
\usepackage{euscript}
\usepackage{tikz}
\usepackage{url}
\usetikzlibrary{topaths,calc}
\usepackage{algorithm,algorithmic}
\usepackage{subfig}
\usepackage{enumerate}

\newtheorem{theo}{Theorem}[section]
\newtheorem{defi}[theo]{Definition}
\newtheorem{prop}[theo]{Proposition}

\DeclareMathOperator{\Ker}{Ker}
\DeclareMathOperator{\im}{Im}

\newcommand{\C}{\mathbb{C}}

\newcommand{\F}{\mathbb{F}}

\newcommand{\h}{\mathcal{H}}

\newcommand{\E}{{\EuScript E}}
\newcommand{\e}{{\EuScript E}}

\newcommand{\Prob}{{\mathbb P}}

\newcommand\ket[1]{|#1\rangle}

\title{A linear-time benchmarking tool for generalized surface codes}

\makeatletter
\let\@fnsymbol\@arabic
\makeatother

\author{Nicolas Delfosse\thanks{IQIM, California Institute of Technology, Pasadena, CA, USA}
$^{,}$\thanks{Department of Physics and Astronomy, University of California, Riverside, CA, USA}\ \ 
Pavithran Iyer\thanks{Département de Physique and Institut Quantique, Université de Sherbrooke, Québec, Canada}
\ and David Poulin\footnotemark[3]}

\begin{document}

\maketitle


\begin{abstract}
Quantum information processors need to be protected against errors and faults. One of the most widely considered
fault-tolerant architecture is based on surface codes. 
While the general principles of these codes are well understood and basic code properties such as minimum distance 
and rate are easy to characterize, a code's average performance depends on the detailed geometric layout of the qubits. 
To date, optimizing a surface code architecture and comparing different geometric layouts relies on costly numerical 
simulations.
Here, we propose a benchmarking algorithm for simulating the performance of 
surface codes, and generalizations thereof, that runs in linear time. 
We implemented this algorithm in a software that generates performance reports
and allows to quickly compare different architectures.
\end{abstract}

\let\thefootnote\relax\footnote{Corresponding author: Nicolas Delfosse - ndelfoss@caltech.edu}

\section{Introduction}
If topological codes are good candidates to support the architecture of a fault-tolerant 
quantum computer, the exact design of this architecture remains to be determined. 
Comparing the performance of different quantum computing architectures is
a costly and time-consuming task that necessitates extensive numerics and is 
often restricted to relatively small system sizes, smaller than what is of practical interest.
For instance, for generic stabilizer codes, verifying whether a residual Pauli error 
after correction is harmful or not requires $O(nk)$ operations where $n$ is the number of physical 
qubits and $k$ is the number of logical qubits. 
For surface codes, the standard decoding algorithm \cite{DKLP02, Ed65a, Ed65b} requires $O(n^3)$ operations.
These costs are further increased when the noise model also incorporates gate and measurement errors, along with the associated error propagation. For instance, the numerical study carried in \cite{BH13:cubic_code} consumed 1000 days of CPU time and those of \cite{NFHDV16} cost 100 000 days of CPU time.

\medskip
Here, we propose a numerical tool to quickly assess and compare the efficiency of different surface code layouts.
The basic idea is to consider a simplified noise model which provide an insight on the performance of quantum error correcting codes at a low cost. Though simplified, the noise model is realistic and provide insight for
fault-tolerant quantum computing due to its similarities with standard noise models
such as the depolarizing channel.
Roughly speaking, this strategy allows us to quickly discriminate between very good codes and bad codes.
Once the best candidates have been identified, more expensive numerics can 
be realized to select the best structure in a restricted list of candidates.

\medskip
The simplified noised model considered in this work is the quantum erasure channel.
Despite its simplicity, the decoding problem for stabilizer codes over this channel
corresponds to solving a linear system which has in general a cubic complexity. 
Achieving a linear time benchmarking is therefore non-trivial. We obtained this result for 
generalized surface codes by studying a notion of induced homology. Roughly, to 
determine if an erasure is correctable, we need to know if the set of erased qubits 
covers a logical error, that is a non-trivial cycle. This is done by computing the rank 
of the homology group covered by the erased edges. If this rank is trivial then 
the erasure can be corrected and we do not need to run a decoder to learn this.

\medskip
Our method can be applied to any surface code, irrespectively of the genus, with or 
without boundaries and where boundaries can be of any type, open, closed or partially open and 
partially closed \cite{DIP16}. We apply this benchmarking tool to two extreme cases.
First, we consider hyperbolic codes. These surface codes
are of particular interest for their constant rate, i.e., they require only a constant number 
of physical qubits per logical qubit. Monte-Carlo simulation of their performance is especially
hard due to the fact that they exhibit important finite-size effects. 
Our numerical tool is able to easily simulate the performance of hyperbolic codes up to length 
larger than $n=100 000$, encoding more than $k=10 000$ qubits. In contrast, previous numerical 
studies of hyperbolic codes were limited to about $1000$ physical qubits in the case of Pauli noise \cite{BT15:hyperbolic}.
Second, we consider the performance of planar code architectures resulting from different 
layouts for the physical qubits. This is motivated by recent and ongoing experimental progress 
towards the realization of these architectures. We propose a software that generates performance 
reports for planar architectures and allows to compare different families of codes.
Our software is available online:

\bigskip
\url{http://quantum-squab.com}

\bigskip
We also note that historically, the erasure channel has play a central role in the development of 
classical error-correcting codes. It is through the study of the erasure channel, and in 
particular the design of decoding algorithms which perform optimally under erasures, 
that capacity achieving and efficiently decodable codes were designed \cite{RU08}.

\medskip
This article is organized as follows. The properties of generalized surface codes are 
recalled in Section~\ref{sec:GSC} and the noise model is described in 
Section~\ref{sec:noise}.
We review the decoding problem for the quantum erasure channel in 
Section~\ref{sec:correctable_erasures}. Our main result is a simple homological
characterization of correctable erasures which is stated in 
Section~\ref{sec:homological_characterization} and proven in Section~\ref{sec:induced_homology} 
through the study of a notion of induced homology. 
Relying on this characterization of correctable erasures, Section~\ref{sec:bench}
proposes a fast benchmarking algorithm for generalized surface codes and Section~\ref{sec:app}
presents two aplications.

\section{Generalized surface codes} \label{sec:GSC}

Surface codes were introduced by Kitaev \cite{Ki03} as
quantum error correcting codes defined by local constraints on qubits placed
on a closed surface, {\em i.e.} a surface without boundaries.
Locality makes surface codes particularly appealing for fault-tolerant
quantum computing \cite{DKLP02, RH07, RHG06, RHG07, FMMC12} as local measurements are sufficient to detect and correct 
errors corrupting encoded states. Moreover, a large class of gates can be 
applied to encoded states in an intrinsically robust (topological or transversal) manner.

\medskip
Kitaev's idea was generalized in two different ways which leads to a fully 
planar design, much simpler for practical realizations.
Bravyi and Kitaev defined code on a plane
using two distinct kind of boundaries: open or closed \cite{BK98}.
Freedman and Meyer generalized this construction to punctured surfaces \cite{FM01}.
The formalism of \cite{DIP16} further generalizes Kitaev construction and these
two extensions by authorizing any surface with or without punctures and
where any boundary along these punctures is an alternate sequence of 
open or closed paths.
Our benchmarking algorithm applies to this whole family of surface codes
that we refer to as generalized surface codes.

\begin{figure}[ht]
\centering
\includegraphics[scale=.7]{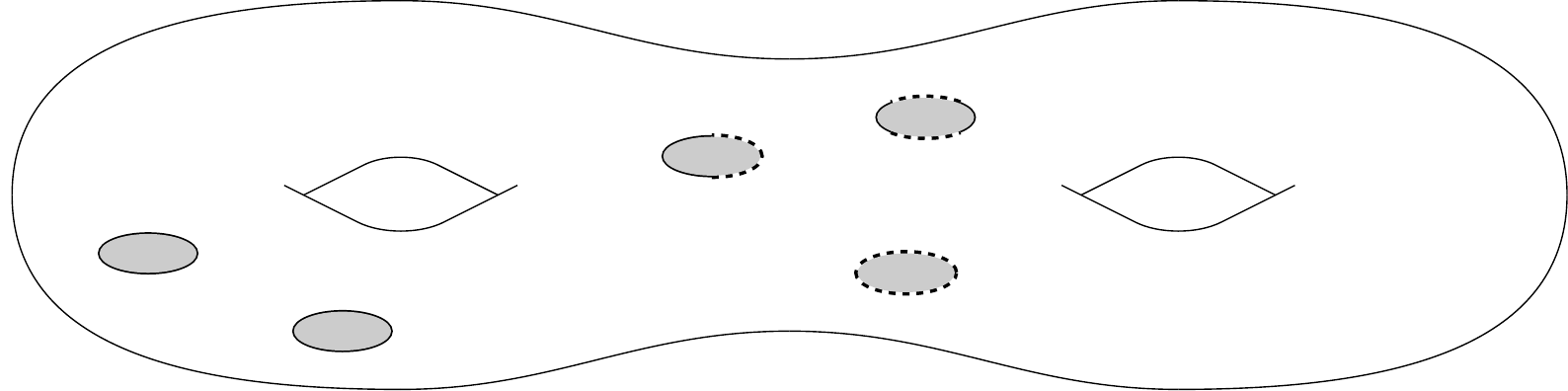}
\caption{A surface of genus 2 (with 2 handles) punctured with 5 holes.
Dotted lines denote open boundaries. }
\label{fig:surface}
\end{figure}

\begin{figure}[ht]
\centering
\includegraphics[scale=.3]{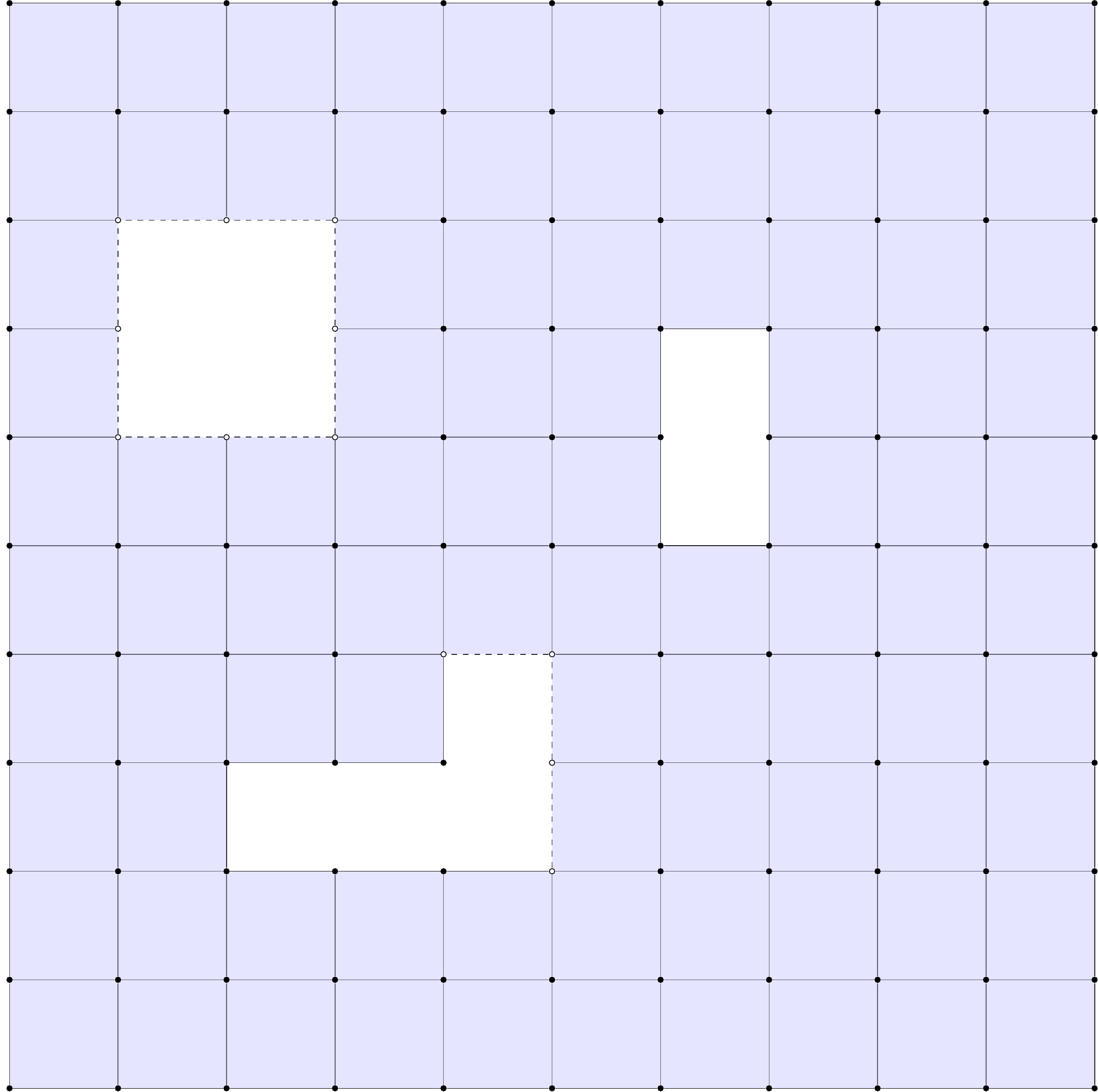}
\caption{A planar square lattice punctured with a closed hole, an open hole
and a hole that is partially closed and partially open.}
\label{fig:surface_lattice}
\end{figure}

\medskip
We now specify the kind of combinatorial surfaces that are used, thereby 
laying out the notation for the rest of this paper.
We consider a cellulation of a surface, that is a surface formed by a set 
of faces glued together along their edges. We denote this combinatorial
surface by the triple $(V, E, F)$ where $V$ is the vertex set, $E$ is the
edge set and $F$ is the face set of the cellulation. We assume that $V$
is a finite set. An edge $e \in E$ is represented as a pair of vertices, 
called its endpoints and a face $f \in F$ is a subset of edges, these edges
are said to be on the boundary of the face $f$.
Some edges  belong to two faces and some edges belong to a single face; the latter form the boundary of the 
surface. The set of boundary edges is denoted by $\partial E$ and their
endpoints are called boundary vertices, denoted by $\partial V$.
Each edge on the boundary will be declared to be either open or closed. 
This induces a partition of the boundary 
$\partial E = \partial_O E \cup \partial_C E$ where
the subscripts $O$ and $C$ refers to open and closed edges.
The two endpoints of an open edge are declared to be open, producing
a similar partition of the boundary vertices 
$\partial V = \partial_O V \cup \partial_C V$.
A special role is played by the set of non-open edges and non-open
vertices that we denote by $\mathring{E}$ and $\mathring{V}$
respectively. Two examples are shown in Figure~\ref{fig:surface}
and Figure~\ref{fig:surface_lattice}.

\medskip
In order to construct a quantum error-correcting code from a combinatorial
surface $G=(V, E, F)$, we place a qubit on each non-open edge of $G$. 
This leads to a Hilbert space $\h = \otimes_{e \in \mathring{E}} \C^2$.
Denote by $X_e$, respectively $Z_e$, the Pauli operator acting as the Pauli
matrix $X$, respectively $Z$, on the qubit indexed by $e$ and as the 
identity on all other qubits.
Then 
\begin{defi} \label{defi:surface_codes}
The {\em surface code} associated with the surface $G=(V, E, F)$ is defined to 
be the ground space of the Hamiltonian
$$
H = - \sum_{v \in \mathring{V}} X_v - \sum_{f \in F} Z_f
$$
where 
$$
X_v = \prod_{\substack{e \in \mathring{E} \\ v \in e}} X_e \quad \text{ and } \quad Z_f = \prod_{\substack{e \in \mathring{E} \\ e \in f}} Z_e.
$$
\end{defi}
When the surface $G$ has no boundaries, this definition coincides with 
Kitaev's original construction. When the surface has only closed boundaries
this is Freedman and Meyer's generalization. Finally, we recover Bravyi and Kitaev's
planar codes when the surface is a sphere with a single puncture.

\medskip
Surface codes are a subclass of stabilizer codes \cite{Go97}. The group generated by 
the operators $X_v$ and $Z_f$ that form the Hamiltonian in Definition~\ref{defi:surface_codes} 
is denoted by $S$ and is called the stabilizer group of the code. 
Any operator of $S$ acts trivially on the ground state of the Hamiltonian, that is on 
encoded states. These operators are called the stabilizers of the code.

\medskip
Error correction for surface codes is based on measurement of the operators
$X_v$ and $Z_f$ defining the Hamiltonian. A non-trivial outcome indicates
the presence of an error since the ground space, {\em i.e.} the code, is fixed
by these operators. When an error $E$ occurs, the outcome of the 
measurement of the stabilizer generators is called the syndrome of the error~$E$.
It is represented as a vector $\sigma(E)$ whose components are the 
measurement outcomes of all the operators $X_v$ and $Z_f$.
Then, the decoder estimates the error which corrupts the state given
its syndrome.

\section{Noise model} \label{sec:noise}

While fault-tolerant architecture are traditionally benchmarked using the depolarizing channel, we choose the erasure channel \cite{GBP97, BDS97} in order to make substantial computational savings. Qubit erasure (or leakage out of the qubit subspace) is often a relevant source of error in different physical realizations . Moreover, performance benchmarks obtained from the erasure channel often share some general trends and features as those obtained from more complex channels. In particular, the erasure channel share many features with the widely studied depolarizing channel.
Roughly speaking, an architecture which performs poorly in the presence of erasures will generally
also have poor performances with other type of errors.
This strategy allows us to quickly identify architectures that are unlikely 
to perform well in a realistic environment at a very low computational cost.

\medskip
Under the action of the quantum erasure channel of erasure probability $p$, each qubit is erased (i.e. lost)
with probability $p$, independently of the other qubits.
When a qubit is erased, it is replaced by a qubit in a
maximally mixed state. Equivalently, we can consider that this qubit is subjected to 
a random Pauli error $I, X, Y$ or $Z$ chosen uniformly. 
In most experimental settings, the location of a lost qubit can be measured easily,
therefore we assume that the decoding algorithm has prior knowledge of the 
location of erased qubits.
This stabilizer description of the quantum erasure channel is well adapted to 
the study of stablizer codes.
It also emphasizes the ressemblance with the depolarizing channel. 
An erased qubit is a completly depolarized qubit, the only difference 
is the knowledge of the erased positions.

\medskip
We will use a $n$-bit vector, $\E \in \F_2^n$, to represent an erasure on 
$n$ qubits, {\em i.e.} the $i$-th component of the vector $\E$ is $\E_i = 1$ 
if the $i$-th qubit is lost and is $\E_i = 0$ otherwise. By definition, the 
probability of a given erasure pattern $\E$ is 
$\Prob(\E) = p^{|\E|} (1-p)^{n-|\E|}$.
Each qubit in the support of $\E$ is corrupted by a random Pauli matrix.
This means that the $n$-qubit state is affected by a random Pauli operator
$E$ chosen uniformly among the $4^{|\E|}$ errors acting non-trivially only 
on the support of $\E$. We denote this condition by $E \subset \E$ and we 
say that the erasure $\E$ covers the Pauli error $E$.
The fact that all Pauli errors which satisfy $E \subset \E$ are equally likely plays a 
crucial role in our benchmarking procedure.

To summarize
\begin{defi} \label{defi:erasure}
An {\em $n$-qubit erasure} is defined to be a pair $(\E, E)$ where 
$\E \in \F_2^n$ is choosen with probability $\Prob(\E) = p^{|\E|} (1-p)^{n-|\E|}$
and $E$ is uniformly distributed among errors $E \subset \E$ covered by $\E$.
\end{defi}

A state $\ket\psi$ which is subjected to such an erasure $(\E, E)$ is mapped onto 
the state $E \ket\psi$. The indicator vector $\E$ is known. The syndrome 
of the error $E$ can be measured.

\medskip
Lastly, we note that our benchmarking method is suited for noise models where 
$\Prob(\E)$ is not independent and identically distributed (i.i.d.), but only consider 
the i..d. case here for simplicity. 

\section{Correction of erasures} \label{sec:correctable_erasures}

We wish to recover a code state $\ket \psi$ which suffers from 
an erasure $(\E, E)$. We assume that the erasure pattern $\E$ is 
known exaclty, as well as the syndrome vector $s = \sigma(E)$.
By definition, errors that belong to the stabilizer group $S$ have no effect 
on encoded states. Ideally, we would like to determine a most likely coset, 
$\tilde E\cdot S$, of an error $\tilde E$ from the knowledge of the 
syndrome $s$ of the error, with the additional information 
$\tilde E \subset \E$. That is a coset $\tilde E\cdot S$ maximizing the
conditional probability $\Prob(\tilde E\cdot S | \E, s)$.
This strategy, called maximum likelihood decoding, is optimal.

\medskip
A striking difference between depolarizing errors and erasures is the 
following dichotomy between erasures that are either easy to correct or 
impossible to correct. Namely, there are two kinds of erasure patterns 
$\E$ which are very different with respect to the maximum likelihood decoding.
The first class is trivial to correct, 
whereas the optimal decoder is bound to fail with a probability at least 
$1/2$ on erasures from the second class (see for instance \cite{DZ13} for a proof).
\begin{prop} \label{prop:erasure_type} 
Any erasure pattern $\E \in \F_2^n$ satisfies one of the two mutually
exclusive properties:
\begin{enumerate}
\item $\E$ does not cover a non-trivial logical error. Then for every syndrome $s$ 
of an error $E \subset \E$, there exists a unique coset $\tilde{E}\cdot S$
with non-zero probability $\Prob(\tilde E\cdot S | \E, s)$ and $\tilde E\cdot S = E\cdot S$.
\item $\E$ covers a non-trivial logical error. Then for every syndrome $s$ of an error $E \subset \E$, 
there exists at least two distinct most likely cosets $\tilde{E}\cdot S$ and $\tilde{E}'\cdot S$.
\end{enumerate}
\end{prop}

An erasure $\E$ satisfying the first condition is said to be {\em correctable}
whereas the second condition defines {\em uncorrectable} erasures.
To correct an erasure $(\E, E)$ of the first type, it suffices to return any estimation
$\tilde E \subset \E$ which has the same syndrome. Then, $\tilde E$
and $E$ have necessarily the same effect on the code, which proves that we 
correctly identified the error which occurs.
On the hand, for an uncorrectable erasure, 
even an optimal decoder is no better than a random guess that fails at least half of
the time.

\section{Characterizing uncorrectable erasures} \label{sec:homological_characterization}

First, we translate the characterization of correctable erasures of 
Proposition~\ref{prop:erasure_type} into a graphical language.
Then, we propose a graphical characterization of uncorrectable erasures,
stated in Theorem~\ref{theo:h1_induced_characterization}
and later proved in Section~\ref{sec:induced_homology}.

\medskip
Recall that a {\em relative cycle} of a surface $G$ is defined to be a subset of 
non-open edges, say $\gamma$,  such that every non-open vertex of $G$ is incident 
to an even number of edges in $\gamma$. 
Few examples of cycles are a closed path, a disjoint union of closed paths but also a path 
connecting two open vertices. A cycle is said to be {\em homologically trivial}, or simply 
{\em trivial}, if it is the non-open boundary of a subset of faces of $G$. 
By non-open boundary, we mean the set of non-open edges on the boundary of a region.
Otherwise such a cycle is called non-trivial.

\medskip
In order to apply Proposition~\ref{prop:erasure_type}, we must identify graphically non-trivial 
logical errors, that is errors of syndrome zero that act non-trivially on the code.
The syndrome of a $Z$-error is given by the measurement of the operators $X_v$ 
associated with non-open vertices. 
These measurements return the parity of the number of $Z$-errors incident to 
a non-open vertex $v$. Therefore, by definition of relative cycles, a $Z$-error 
has trivial syndrome if and only if its support is a relative cycle of $G$.
Only some of these errors have a non-trivial action on encoded states.
For instance, any stabilizer generator $Z_f$, whose support is the boundary of the face 
$f$ has syndrome zero but it also acts trivially.
So does any $Z$-stabilizer which is supported on the boundary of a region.
Logical errors acting non-trivially on encoded states correspond to non-trivial cycles of the graph \cite{DIP16}.
A similar argument can be made for $X$-errors, but using the dual graph. 

\begin{prop} \label{prop:correctable_equiv_no_homology}
An erasure $(\E, E)$ is correctable if and only if $\E$ does not cover a 
non-trivial cycle in the tiling $G$ or in its dual $G^*$.
\end{prop}

This proposition is the combined conclusion of the desciption of logical 
operators for generalized surface codes \cite{DIP16} and 
Prop.~\ref{prop:erasure_type}.
The notion of dual considered in this article is the generalized dual 
introduced in \cite{DIP16}.

\medskip
We must therefore find a strategy to determine if an erased pattern $\E$ 
covers a non-trivial homology in the graph G.
A similar argument will then be applied to its dual.
Our main result is Theorem\ref{theo:h1_induced_characterization} on which we rely to determine 
the number of homologically non-trivial cycles covered by an erasure $\E$
in a surface $G$. If this number is non-zero, the erasure covers at least one
non-trivial cycle and is thus uncorrectable.

\medskip
In what follows, $G^*$ is the dual cellulation of $G^*$.
For $\e \subset \mathring{E}$, denote by $\bar \e$ the complementary 
of the set $\e$ in $\mathring{E}$. By duality, this set can also be regarded as a subset 
of edges of the dual graph $G^*$. 
Then define the graph $G^*_{\bar \e}$ as the subgraph of $G^*$ of vertex set 
$V^*$ and edge set $\bar \e$.
The notation $\kappa_{X}(H)$ (respectively $\kappa_{\bar X}(H)$) refers
to the number of connected components of the graph $H$ containing at
least one element of the set $X$ (respectively no elements of the set $X$).
For instance, $\kappa_{\overline{\partial_O V}}(H)$ is the number of connected 
components of $H$ containing no open vertex.
The graph $H_{\e}$ is the subgraph of $H = (V, E)$ with the same vertex
set $V$ but with edge set $\e$. 

\medskip
In order to state our main result which provides a characterization of correctable 
erasures, let us introduce the following notation
$$
h_1(G, \E) =  |\e| - |\mathring{V}| + \kappa_{\overline{\partial_O V}}(G_\e) - \kappa_{\overline{\partial_O V^*}}(G^*_{\bar \e}) + \kappa_{\overline{\partial_C E}}(G)
$$
where $G$ is a surface and $\E \subset \mathring E$ is a subset of non-open
edges of $G$.
We will also make use of $h_1(G^*, \E^*)$ where $G^*$ is the dual cellulation and 
$\E^*$ is the subset of non-open egdes of the dual graph corresponding to 
the edges of $\E \subset \mathring E$.

\begin{theo} \label{theo:h1_induced_characterization}
Consider a tiling $G$ and a subset $\e \subset \mathring{E}$ of its edge set.
The erasure $\E$ is correctable if and only if $h_1(G, \E) + h_1(G^*, \E^*) = 0$.
\end{theo}

This formula may appear complicated, however it only involves quantities that can be computed 
in linear time. Indeed, finding the number of vertices, edges
or the number of connected components of a graph has a linear cost.
More precisely, $h_1(G, \E) > 0$ if and only $\E$ covers a logical $Z$-error 
whereas $h_1(G^*, \E^*) > 0$ detects logical $X$-errors covered by $\E$.
We can therefore also analyze the performance of the code to correct only 
$X$-errors or only $Z$-errors.

\medskip
The number $h_1(G, \E)$ is the dimension of the homology group $H_1(G_\e)$ 
covered by the edges of $\E$. This group is formally defined in 
Section~\ref{sec:induced_homology} where Theorem~\ref{theo:h1_induced_characterization}
is proven.
Roughly speaking, $h_1(G, \E)$ counts the number of independent 
non-trivial cycles covered by the erasure $\E$.

\section{Induced homology} \label{sec:induced_homology}

The purpose of this section is to prove Theorem~\ref{theo:h1_induced_characterization}.
Our basic idea is to study the homology covered by an erasure $\e$.

\medskip
First let us recall the definition of the chain complex 
$
C_2 \overset{\partial_2}{\longrightarrow} C_1 \overset{\partial_1}{\longrightarrow} C_0
$
defining homology in a  surface $G = (V, E, F)$ with open and closed boundaries.
We use the notation of \cite{DIP16}. For a more complete reference about 
homology see for instance \cite{Ha02, Gi10}.
This chain complex is a triple of $\F_2$-linear spaces equipped with 2 $\F_2$-linear maps.
The spaces $C_0, C_1, C_2$ are respectively the formal sum of non-open vertices, 
non-open edges and faces with binary coefficients, that is
$$
C_0 = \bigoplus_{v \in \mathring{V}} \F_2 v, \quad C_1 = \bigoplus_{e \in \mathring{E}} \F_2 e, \quad C_2 = \bigoplus_{f \in F} \F_2 f \cdot
$$
For instance a vector of $C_2$ is a sum $\sum_{f \in F} \lambda_f f$ with $\lambda_f \in \F_2$.
The map $\partial_2$ associates with a face $f$ the sum of the non-open edges living on its boundary
and $\partial_1$ associates with an edge the sum of its non-open endpoints. 
For instance, if $f = \{e_1, \dots, e_m\}$ is a face and if $e_m$ is its only open edge then
$\partial_2(f) = \sum_{i=1}^{m-1} e_i \in C_1$.
%
Let us now connect this formalism with the definition of relative cycles and
homologically trivial cycles introduced in the previous section.
Any vector $x$ of $C_1$ is the indicator vector of a subset $\gamma_x$ 
of non-open edges of $G$. 
We can easily see that this subset $\gamma_x$ is a relative cycle if and only if
$x$ is in the kernel of the map $\partial_1$. Moreover, it is homologically trivial 
if and only if $x$ belongs to the image of the map $\partial_2$.
Henceforth, we will denote the support of an error by one of its two equivalent forms 
-- a subset of non-open edges of $G$, or a vector of $C_1$, depending upon the context.
The size of the set of non-trivial cycles can be measured by the size of the first homology 
group with binary coefficients
$
H_1(G) = \Ker \partial_1 / \im \partial_2.
$
It is the set of relative cycles up to boundaries. Cycles whose coset is non-trivial in this quotient
are exactly non-trivial cycles.
This quotient is well defined since the composition of the 2 boundary maps is trivial.
Our goal is to introduce a generalization of this homology group that would count the number
of non-trivial cycles covered by a subset of edges $\e \subset \mathring E$.

\medskip
In order to define the homology group induced by the subgraph $G_\e$, we 
introduce the following chain spaces
$$
C_0^\e=C_0, \quad C_1^\e = \bigoplus_{e \in \e} \F_2 e,
\quad C_2^\e = \partial_2^{-1}(C_1^\e).
$$
equipped with the restrictions $\partial_1^\e$ and $\partial_2^\e$ of 
$\partial_1$ and $\partial_2$ to $C_1^\e$ and $C_2^\e$.
By construction, these transformation inherit of the structure of chain complex
of $\partial_2, \partial_1$, {\em i.e.} they satisfy 
$\partial_1^\e \circ \partial_2^\e = 0$. Subsequently, we have 
$\im \partial_2^\e \subset \Ker \partial_1^\e$, allowing to define the 
induced homology group as follows.

\begin{defi} \label{defi:induced _homology_group}
Let $\e \subset \mathring{E}$ be a subset of the non-open edges of a surface $G$. 
The \emph{first homology group induced by $\e$},
denoted by $H_1(G_{\e})$, is the quotient space
$$
H_1(G_{\e}) = \Ker \partial_1^{\e} / \im \partial_2^\e.
$$
\end{defi}

By definition, the elements of $\Ker \partial_1^{\e}$ are the cycles of $G$
that are included in the subset $\e$ and elements of $\im \partial_2^\e$
are the boundaries or trivial cycles included in $\e$.
This quotient is trivial if and only if every cycle of $G$ covered by $\E$ is
a trivial cycle.

\medskip
The induced homology group $H_1(G_{\e})$ considered here, just like the homology group
of the surface, is equipped with a structure of $\F_2$-linear space.
The following proposition allows us to compute the dimension of this group. 
Theorem~\ref{theo:h1_induced_characterization} is an immediate application 
of this proposition. Namely, there exists a non-trivial cycle covered by the erasure $\e$
if and only if the group $H_1(G_\e)$ is non-trivial, that is if and only if its dimension
is non-zero.

\begin{prop} \label{prop:dim_H_1_induced}
Let $\e \subset \mathring{E}$ be a subset of the non-open edges of a surface $G$. 
Then, we have
$$
\dim H_1(G_{\e}) = |\e| - |\mathring{V}| + \kappa_{\overline{\partial_O V}}(G_\e) - \kappa_{\overline{\partial_O V^*}}(G^*_{\bar \e}) + \kappa_{\overline{\partial_C E}}(G) \cdot
$$
\end{prop}

Recall that $\kappa_{X}(H)$ (respectively $\kappa_{\bar X}(H)$) was defined
before Theorem~\ref{theo:h1_induced_characterization} as the number of 
connected components of the graph $H$ containing at least one element of 
the set $X$ (respectively no elements of the set $X$).

\medskip
For a connected closed surface that is orientable of genus $g$ with $\E = E$, 
we recover the standard formula 
$
\dim H_1(G_{\E}) = - |V| + |E| - |F| + 2 = 2g.
$
Therein, we used $\kappa_{\overline{\partial_O V^*}}(G^*_{\bar \e}) = |F|$,
and $\kappa_{\overline{\partial_O V}}(G_\e) = \kappa_{\overline{\partial_C E}}(G) = 1$.
At the other extreme, when $\e = \emptyset$, we find
$
\dim H_1(G_{\e}) 
= 0
$
as expected.

\begin{proof}
By definition, the dimension of $H_1(G_{\e})$ is given by 
$\dim \Ker \partial_1^{\e} - \dim \im \partial_2^\e$.
The space $\Ker \partial_1^{\e}$ is the cycle space of the 
open graph $G_\e = (V, \e)$ with open-vertex set $\partial_O V$.
Its dimension was computed in Proposition~4.1 of \cite{DIP16}:
\begin{equation} \label{eqn:dim_ker_partial_2_e}
\dim \Ker \partial_1^{\e} = |\e| - |\mathring{V}| + \kappa_{\overline{\partial_O V}}(G_\e).
\end{equation}
Our next task is to determine the dimension of the image of the map 
$\partial_2^\e: C_2^\e \rightarrow C_1^\e$. We first compute the 
dimension the space $C_2^\e$.

\medskip
\noindent
{\em Step 1. Dimension of $C_2^\e$:}
First, let us describe the vectors of $C_2^\e$. 
By definition, a sum 
$\sum_{f \in I} f$, where $I \subset F$, belongs to $C_2^\e$ if and only 
if the vector $\partial_2(\sum_{f \in I} f)$ of $C_1$ is the indicator vector 
of a set of edges included in $\e$, \emph{i.e.} a vector of the form 
$\sum_{e \in A} e$ where $A$ is a subset of $\e$. Using 
$\partial_2(\sum_{f \in I} f) = \sum_{f \in I} \partial_2(f)$,
we obtain the charaterization: $\sum_{f \in I} f \in C_2^\e$ if and only if

\medskip
(i) any edge of the set $\bar \e = \mathring{E} \backslash \e$ appears 
on the boundary $\partial_2(f)$ of 0 or 2 faces $f$ of $I$.
\medskip

\noindent
To simplify this condition let us introduce the subgraph 
$G_{\bar \e}^* = (V^*, \bar \e)$ of $G^*$ induced
by the edges of $\bar \e$. 
The set $I \subset F$ corresponds to the subset 
$
I^* = \{ v_f \in V^* \ | \ f \in I \}
$
of the vertex set of $G^*_{\bar \e}$.
Then criterion (i) takes the form $\sum_{f \in I} f \in C_2^\e$ if and only if

\medskip
(ii) any edge of $\bar \e$ is incident to either 0 or 2 vertices of $I^*$
in the graph $G^*_{\bar \e}$.
\medskip

\noindent
This implies that whenever $I^*$ contains a vertex $v$ of $G^*_{\bar \e}$, 
it contains all its neighbors in $G^*_{\bar \e}$, but also the neighbors of
its neighbors. Therefore, in order to respect (ii), 
$I^*$ must contain the whole connected component of the vertex $v$.
Moreover, since $I^*$ is composed exclusively of non-open vertices of 
$G^*_{\bar \e}$, if the connected components of $v$ satisfies (ii), it 
cannot contain any open vertex. Conversely, such a connected component 
clearly satifies (ii).
This leads to the new characterization: $\sum_{f \in I} f \in C_2^\e$ if and only if

\medskip
(iii) $I^*$ is a union of connected components of the graph $G^*_{\bar \e}$ 
containing no open vertices.
\medskip

\noindent
Given $\e$, write $(C^*_j)_{j \in J}$, for some finite set $J$, 
the set of connected components of $G_{\bar \e}^*$ containing no 
open vertices $v \in \partial_O V^*$ and denote by $C_j$ the 
corresponding subsets of faces in $G$. The characterization (iii) proves 
that the vectors of the $C_2^\e$ are 
of the form
\begin{equation} \label{eqn:vectors_of_C_2_e}
\sum_{j \in J} \lambda_j \left( \sum_{f \in C_j} f \right),
\end{equation}
for some family of binary coefficients $\lambda_j \in \F_2$.
Since the vectors $\sum_{f \in C_j} f$ are linearly independent, they
form a basis of the linear space $C_2^\e$. In other words, we just 
showed that the dimension of $C_2^\e$ is
\begin{equation} \label{eqn:dim_C_2_e}
\dim C_2^\e = \kappa_{\overline{\partial_O V^*}}(G^*_{\bar \e})
\end{equation}
which is the number of connected components of the graph $G^*_{\bar \e}$
containing no open vertices.

\medskip
\noindent
{\em Step 2. Dimension of $\im \partial_2^\e$:}
We are now in position to compute the dimension of the kernel and then 
the dimension of the image of the application $\partial_2^\e$. 
For a vector of $C_2^\e$ given by Eq~\eqref{eqn:vectors_of_C_2_e},
being in the Kernel of $\partial_2^\e$ is equivalent to have
\begin{equation} \label{eqn:kernel_partial_2}
\sum_{j \in J} \lambda_j \partial_2^\e \left( \sum_{f \in C_j} f \right) = 0
\end{equation}
in $C_2^\e$.
To simplify, denote by $\partial_2^\e(C_j)$ the term 
$\partial_2^\e \left( \sum_{f \in C_j} f \right)$
which represents the set of non-open edges on the boundary of the 
$j$-th component.

\medskip
Assume first that the graph $G$ is a connected graph with no closed boundaries 
$e \in \partial_C E$, then both $G^*$ and $G^*_{\bar \e}$ have no open vertices. 
Thus all the connected components of $G^*_{\bar \e}$ satisfy criterion~(iii). 
Moreover an edge of $\e$ cannot belong to a single set $\partial_2^\e(C_j)$ 
otherwise it would be a closed-boundary edge. Therefore, we get
$
\sum_{j \in J} \partial_2^\e(C_j) = 0,
$
providing an element $\sum_{j \in J} \sum_{f \in C_j} f$ of the kernel.
This is the only non-trivial vector of $\Ker \partial_2^\e$ since removing 
any component $C_i$ introduces a non-open edge of $\e$ that belongs to 
a single component $C_j$ with $j \neq i$. This forbids another relation of 
the form of Eq.\eqref{eqn:kernel_partial_2}. Hence, the space 
$\ker \partial_2^\e$ has dimension 1 and the dimension of the image 
of $\partial_2^\e$ is given by
$$
\dim \im \partial_2^\e = \kappa_{\overline{\partial_O V^*}}(G^*_{\bar \e}) - 1.
$$

\medskip
We found the dimension of $\ker \partial_2^\e$ assuming that $G$ is connected
and has no closed boundary. Now, consider that $G$ is connected but containing 
an edge $e \in \partial_C E$. Either no components of $G^*_{\bar \e}$ pass 
criterion (iii) or there exists an edge $e \in \e$ that belongs to a single
component satisfying this criterion. In both cases the kernel of $\partial_2^\e$
is trivial and the space $\im \partial_2^\e$ has dimension
$$
\dim \im \partial_2^\e = \kappa_{\overline{\partial_O V^*}}(G^*_{\bar \e}).
$$

\medskip
Applying one of the two possible dimension formulas for $\im \partial_2^\e$ to each of 
the connected components of $G$, we find
\begin{equation} \label{eqn:im_partial_2_e}
\dim \im \partial_2^\e = \kappa_{\overline{\partial_O V^*}}(G^*_{\bar \e}) - \kappa_{\overline{\partial_C E}}(G).
\end{equation}
The theorem follows from Eq.\eqref{eqn:dim_ker_partial_2_e} and \eqref{eqn:im_partial_2_e}.
\end{proof}

\section{Benchmarking algorithm} \label{sec:bench}

The performance of quantum error correcting codes are usually estimated
by Monte-Carlo simulation. In standard error correction benchmarking, the 
following steps are repeated a large number of times.\\

\begin{enumerate}[(i)]
\item\label{item_error} Generate a random error $E$.
\item\label{item_syndrome} Compute the associated syndrome $\sigma(E)$
\item\label{item_decode} Use a decoder to estimate the error from its syndrome.
\item\label{item_check} Check if the estimation is equivalent to the original error up to a stabilizer.
\end{enumerate}

The cost of (\ref{item_decode}) depends on the complexity of the decoder being used. 
For instance, it is $O(n^3)$ for the standard minimum weight perfect matching
decoder for a depolarizing noise. 
In general, decoding over the quantum erasure channel corresponds to solving a linear system
which would lead to a cubic complexity as well.
The complexity of the step (\ref{item_check}) is usually $O(kn)$ where $k$ is the number of logical operators.

\medskip
In our scheme we directly determine whether the optimal decoder will succeed or not 
without actually running the decoder and checking its results.
This allows us to replace steps (\ref{item_syndrome}) to (\ref{item_check}) by a single step which has a linear-time complexity.
Algorithm~\ref{algo:bench} is the main routine of our benchmarking 
algorithm. It takes as an input a combinatorial surface $G$, its dual $G^*$ and an erasure 
$\E$ and it tells us whether this erasure is correctable or not for the 
surface code associated with $G$.
The key ingredient of this algorithm is Theorem~\ref{theo:h1_induced_characterization}.

\begin{algorithm}
\caption{Benchmarking main routine}
\label{algo:bench}

\begin{algorithmic}[1]
\REQUIRE A tiling $G = (\mathring{V}, \mathring{E}, F)$ and its dual $H = G^* = (\mathring{V}^*, \mathring{E}^*, F^*)$, an erasure $\E$.
\ENSURE Determine whether $\E$ is correctable for the surface code based on $G$.

\STATE Compute $h_1(G, \E)$ and $h_1(G^*, \E^*)$ 
\STATE If $h_1(G, \E) + h_1(G^*, \E^*) >0$, return {\bf uncorrectable}.
\STATE Else return {\bf correctable}.

\end{algorithmic}
\end{algorithm}

The computation of the size $|\E|$ of the erasure and the number of vertices 
$|\mathring{V}|$ and $|\mathring{V}^*|$ is obviously possible in linear time.
It is also straightforward to determine the number of connected components 
of a graph in linear time. The conditions given by the subscripts of $\kappa$ do
not make the problem any more difficult.
The complexity of Algorithm~\ref{algo:bench} is linear.

\section{Two applications of our benchmarking algorithm} \label{sec:app}
\subsection{Performance of hyperbolic codes}

Our first application illustrates the efficiency of our algorithm, by simulating 
codes with a very large number of qubits.
We consider hyperbolic surface codes \cite{FML02, Ki07, Ze09, DZ10, Del13:tradeoffs, BT15:hyperbolic}. 
These codes are defined from tilings of manifolds of
large genus and have the advantage of encoding a large number of qubits.
Namely, encoding $k$ qubits with a growing minimum distance $d$ requires 
only $n = C k$ physical qubits for some constant $C$. This induces a much lower 
overhead than with surface codes on square lattices.

These codes are known to have a threshold for depolarizing errors \cite{KP12},
however this property has never been observed numerically since it requires us to 
consider codes over a very large number of physical qubits.

\begin{figure}[ht]
\hspace{-1.5cm}
\includegraphics[scale=.38]{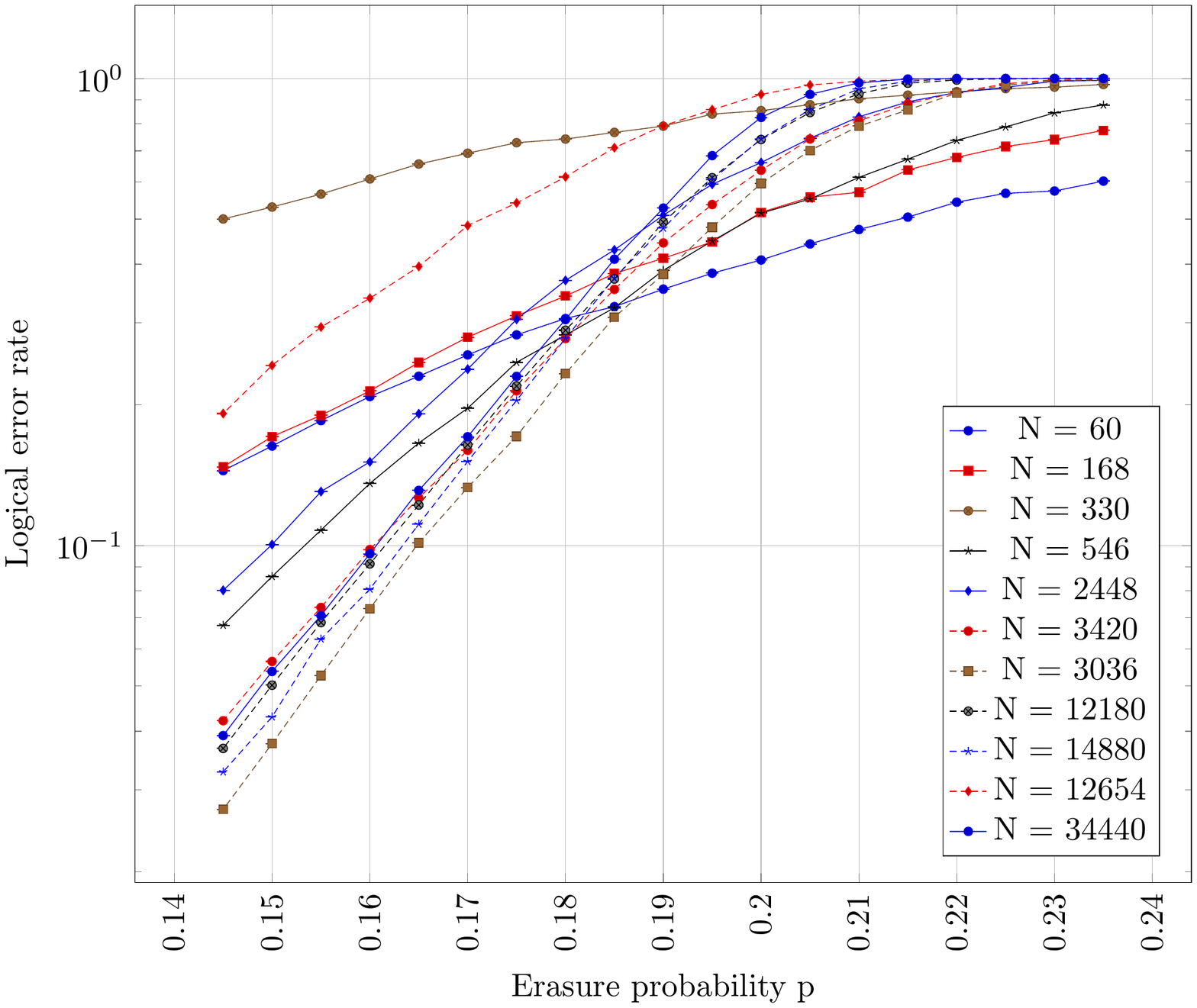}
\hspace{-2.5cm}
\includegraphics[scale=.38]{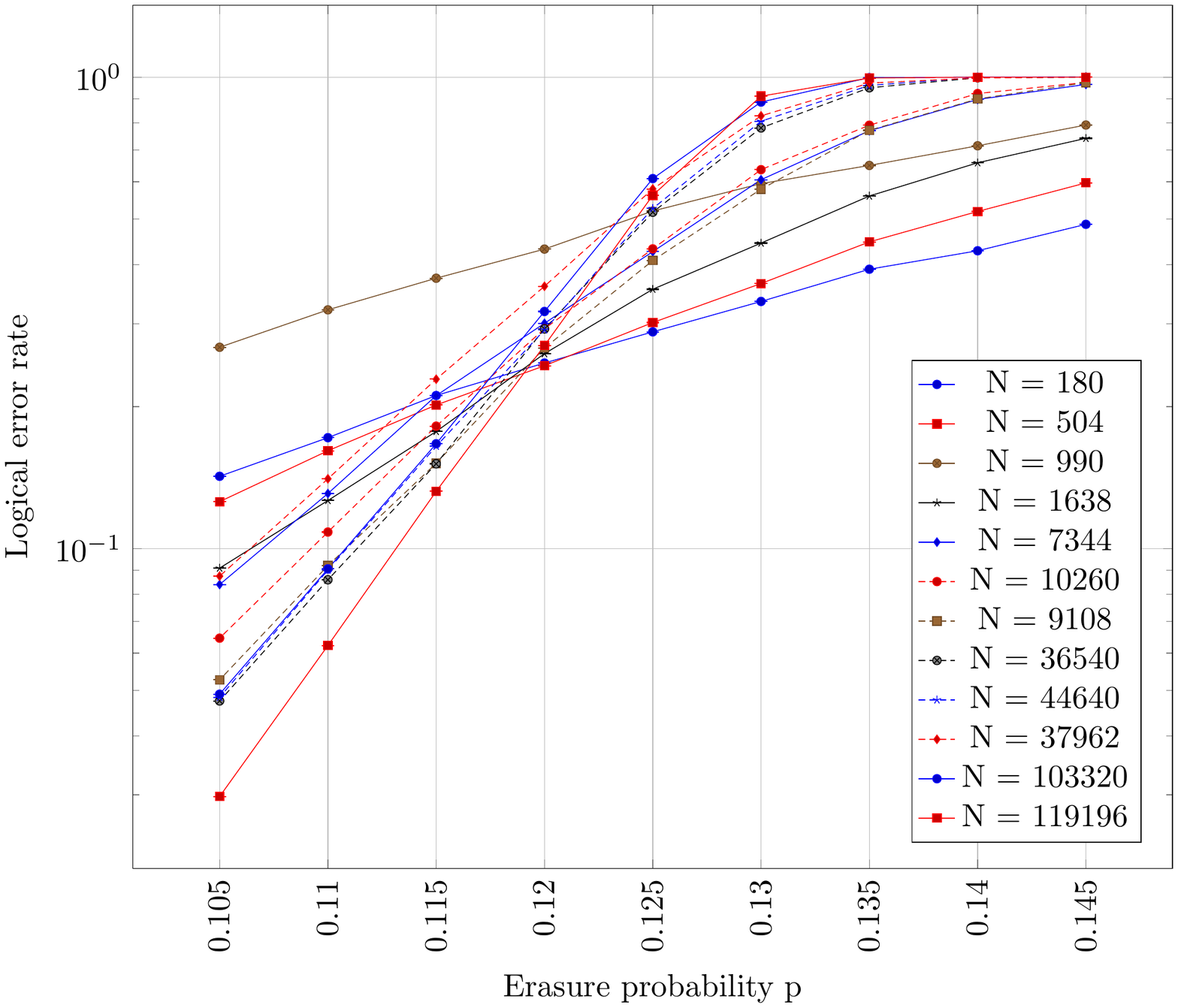}
\vspace{-1cm}
\caption{Performance of two families of hyperbolic codes of varying length $n$.
Left: Based on a (6,6) self-dual lattice with rate $k/n\geq 1/3$ \cite{DZ10}.
Right: Based on a trivalent lattices with faces of length 6 and 12 with rate $k/n \geq 1/9$ \cite{Ze09}.
Generating the data took less than 1 day of CPU time.}
\label{fig:hyperbolic}
\end{figure}

Using our benchmarking algorithm, we are able to simulate the performance of 
hyperbolic codes of length up to $n=100 000$ encoding more than $k=10 000$
qubits and we observe a behavior that suggest a threshold for these codes.
Note that these data were obtained in less than 1 day of CPU time.
These perfomance are depicted in Figure~\ref{fig:hyperbolic}.
The first family is based on (6,6) lattices, that is lattices with degree 6 vertices and 
hexagon faces, and was constructed in \cite{DZ10}. Our second example is made from
3-regular lattices with 2 types of faces, having lengh 6 and 12 \cite{Ze09}.
The performance of similar codes was simulated by standard Monte-Carlo simulation 
for length up to $n=1200$ in \cite{BT15:hyperbolic}, and this size was too small to 
observe a clear threshold.

\subsection{Benchmarking software for planar quantum computing architectures}

Based on Algorithm~\ref{algo:bench}, we implemented a software 
which estimates the performance of surface codes by homological benchmarking 
(Theorem~\ref{theo:h1_induced_characterization}).
This software allows to
\begin{enumerate}
\item Construct and display any planar lattice with or without holes and with open and closed boundaries.
\item Plot the performance of corresponding surface code estimated by homological benchmarking.
\item Compare the performance of different codes by homological benchmarking.
\end{enumerate}
The performance report generated contains the parameters of the code, the weight distribution of the 
measurement required, the performance for correction of $X$-error, $Z$-error and for correction of both 
types of error simultaneously.
Our software, illustrated with different examples, can be downloaded online on the website:

\bigskip
\url{http://quantum-squab.com}

\section{Concluding remarks}

We proposed a fast algorithm to estimate the performance of generalized surface codes.
Our strategy allows us to quickly analyze the robustness of a given layout of qubits, helping us
to optimize quantum computing architectures based on surface codes.
Note that we are able to probe the optimal performance of a surface code for error-correction
without actually runing a decoding algorithm. 
The homological tools developped here are actually not sufficient to design a decoder. 
We can determine whether or not an erasure is correctable by computing the number
of logical errors covered by a given erasure but we cannot identify the error which occurs.
In an upcoming work, a linear-time decoder for correcting erasures with generalized surface 
codes is proposed. 
A decoding algorithm for correction of erasures, based on measurement superplaquettes,
obtained by fusioning faces of the tiling, was proposed by Barrett and 
Stace \cite{barrett2010:loss}. Beyond erasure, it allows to correct combination for Pauli 
errors and erasures but its complexity is clearly super-linear.

\medskip
For future work, more complicated noise model could be considered.
For instance, simulating the performance of surface codes against correlated erasures
can also be done in linear time with the same approach.
We may also extend our strategy to other families of codes. Natural candidates 
for this generalization are color codes \cite{BM06:color}
or surface codes with twist defects \cite{bombin2010:twist}.
Twists enable a more direct fault-tolerant implementation of certain gates \cite{brown2016:hybrid_twist_hole}.

\medskip
More generally, one may wonder if this benchmarking tool can be adapted
to quantum Low Density Parity--Check (LDPC) codes \cite{MMM04}.
In our quest for a good fault-tolerant architecture for quantum computing,
LDPC codes are one of the most serious chalenger for surface codes.
Their main advantage over planar surface codes is that they could considerably
reduce the overhead \cite{gottesman2014:LDPC}. Only a constant number of physical qubit per logical qubit 
would be sufficient to ensure fault-tolerance. However, an indeepth analysis of the
performance of quantum LDPC code is missing. One of the main reason is that a good
decoder is missing in general \cite{MMM04, PC08, delfosse2014:correlations, leverrier2015:expander}.
Our benchamrking approach, which does not even require a decoding algorithm, 
provides an ideal framework for such a study, athough preserving 
a linear cost seems difficult in general. We leave this question open for future work.

\section{Acknowlegdments}
The authors would like to thank Marcus da Silva for enlighting discussions
and Mario Berta and Tomas Jochym-O'Connor for their comments on a 
preliminary version of our software.
ND acknowledges funding provided by the Institute for Quantum Information and Matter, 
an NSF Physics Frontiers Center (NSF Grant PHY-1125565) with support of 
the Gordon and Betty Moore Foundation (GBMF-2644).
This work was supported by the Army Research Office contract number
W911NF-14-C-0048. PI and DP are supported by Canada's NSERC and by the 
Canadian Institute for Advanced Research.

%

\begin{thebibliography}{10}

\bibitem{DKLP02}
E.~Dennis, A.~Kitaev, A.~Landahl, and J.~Preskill.
\newblock Topological quantum memory.
\newblock {\em Journal of Mathematical Physics}, 43:4452, 2002.

\bibitem{Ed65a}
J.~Edmonds.
\newblock Maximum matching and a polyhedron with 0-1 vertices.
\newblock {\em Journal of Research at the National Bureau of Standards},
  69B:125--130, 1965.

\bibitem{Ed65b}
J.~Edmonds.
\newblock Path, trees, and flowers.
\newblock {\em Canadian Journal of Mathematics}, 17:449--467, 1965.

\bibitem{BH13:cubic_code}
S. Bravyi and J. Haah.
\newblock Quantum self-correction in the 3d cubic code model.
\newblock {\em Phys. Rev. Lett.}, 111:200501, Nov 2013.

\bibitem{NFHDV16}
S. Nagayama, A.~G Fowler, D. Horsman, S.~J Devitt, and R.
  Van~Meter.
\newblock Surface code error correction on a defective lattice.
\newblock {\em arXiv preprint arXiv:1607.00627}, 2016.

\bibitem{DIP16}
N. Delfosse, P. Iyer, and D. Poulin.
\newblock Generalized surface codes and packing of logical qubits.
\newblock {\em arXiv preprint arXiv:1606.07116}, 2016.

\bibitem{BT15:hyperbolic}
N.~P Breuckmann and B.~M Terhal.
\newblock Constructions and noise threshold of hyperbolic surface codes.
\newblock {\em arXiv preprint arXiv:1506.04029}, 2015.

\bibitem{RU08}
T.~Richardson and R.~Urbanke.
\newblock {\em Modern Coding Theory}.
\newblock Cambridge University Press, 1 edition, 2008.

\bibitem{Ki03}
A.Y. Kitaev.
\newblock Fault-tolerant quantum computation by anyons.
\newblock {\em Annals of Physics}, 303(1):27, 2003.

\bibitem{RH07}
R.~Raussendorf and J.~Harrington.
\newblock Fault-tolerant quantum computation with high threshold in two
  dimensions.
\newblock {\em Physical Review Letters}, 98(19):190504, 2007.

\bibitem{RHG06}
R.~Raussendorf, J.~Harrington, and K.~Goyal.
\newblock A fault-tolerant one-way quantum computer.
\newblock {\em Annals of Physics}, 321(9):2242 -- 2270, 2006.

\bibitem{RHG07}
R.~Raussendorf, J.~Harrington, and K.~Goyal.
\newblock Topological fault-tolerance in cluster state quantum computation.
\newblock {\em New Journal of Physics}, 9:199, 2007.

\bibitem{FMMC12}
A.~G. Fowler, M.~Mariantoni, J.~M Martinis, and A.~N Cleland.
\newblock Surface codes: Towards practical large-scale quantum computation.
\newblock {\em Physical Review A}, 86(3):032324, 2012.

\bibitem{BK98}
S.B. Bravyi and A.Y. Kitaev.
\newblock Quantum codes on a lattice with boundary.
\newblock arXiv:9811052, 1998.

\bibitem{FM01}
M.~H. Freedman and D.~A. Meyer.
\newblock Projective plane and planar quantum codes.
\newblock {\em Foundations of Computational Mathematics}, 1(3):325--332, 2001.

\bibitem{Go97}
D.~Gottesman.
\newblock {\em Stabilizer Codes and Quantum Error Correction}.
\newblock PhD thesis, California Institute of Technology, 1997.

\bibitem{GBP97}
M.~Grassl, T.~Beth, and T.~Pellizzari.
\newblock Codes for the quantum erasure channel.
\newblock {\em Physical Review A}, 56:33--38, 1997.

\bibitem{BDS97}
C.H. Bennett, D.P. DiVincenzo, and J.A. Smolin.
\newblock Capacities of quantum erasure channels.
\newblock {\em Physical Review Letters}, 78:3217--3220, 1997.

\bibitem{DZ13}
N. Delfosse and G. Z{\'e}mor.
\newblock Upper bounds on the rate of low density stabilizer codes for the
  quantum erasure channel.
\newblock {\em Quantum Information \& Computation}, 13(9-10):793--826, 2013.

\bibitem{Ha02}
A.~Hatcher.
\newblock {\em Algebraic topology}.
\newblock Cambridge University Press, 2002.

\bibitem{Gi10}
P.~Giblin.
\newblock {\em Graphs, surfaces and homology}.
\newblock Cambridge University Press, 2010.

\bibitem{FML02}
M.H. Freedman, D.A. Meyer, and F.~Luo.
\newblock Z2-systolic freedom and quantum codes.
\newblock {\em Mathematics of Quantum Computation, Chapman \& Hall/CRC}, pages
  287--320, 2002.

\bibitem{Ki07}
I.H. Kim.
\newblock {\em Quantum codes on Hurwitz surfaces}.
\newblock PhD thesis, Massachusetts Institute of Technology, 2007.

\bibitem{Ze09}
G.~Z{\'e}mor.
\newblock On {C}ayley graphs, surface codes, and the limits of homological
  coding for quantum error correction.
\newblock In {\em Proc. of the 2nd International Workshop on Coding and
  Cryptology, IWCC 2009}, pages 259--273. Springer-Verlag, 2009.

\bibitem{DZ10}
N.~Delfosse and G.~Z\'emor.
\newblock Quantum erasure-correcting codes and percolation on regular tilings
  of the hyperbolic plane.
\newblock In {\em Proc. of IEEE Information Theory Workshop, ITW 2010}, pages
  1--5, 2010.

\bibitem{Del13:tradeoffs}
N. Delfosse.
\newblock Tradeoffs for reliable quantum information storage in surface codes
  and color codes.
\newblock In {\em Proc. of IEEE International Symposium on Information Theory,
  ISIT 2013}, pages 917--921, 2013.

\bibitem{KP12}
A.~A. Kovalev and L.~P. Pryadko.
\newblock Fault tolerance of quantum low-density parity check codes with
  sublinear distance scaling.
\newblock {\em Phys. Rev. A}, 87:020304, Feb 2013.

\bibitem{barrett2010:loss}
S.~D Barrett and T.~M Stace.
\newblock Fault tolerant quantum computation with very high threshold for loss
  errors.
\newblock {\em Physical review letters}, 105(20):200502, 2010.

\bibitem{BM06:color}
H.~Bombin and M.A. Martin-Delgado.
\newblock Topological quantum distillation.
\newblock {\em Physical Review Letters}, 97:180501, 2006.

\bibitem{bombin2010:twist}
H.~Bombin.
\newblock Topological order with a twist: Ising anyons from an abelian model.
\newblock {\em Physical review letters}, 105(3):030403, 2010.

\bibitem{brown2016:hybrid_twist_hole}
M. S.~Kesselring B. J.~Brown, K.~Laubscher and J.~R.
  Wootton.
\newblock Poking holes and cutting corners to achieve clifford gates with the
  surface code.
\newblock arXiv:1609.04673, 2016.

\bibitem{MMM04}
D.~J.~C. MacKay, G.~Mitchison, and P.~L. McFadden.
\newblock Sparse-graph codes for quantum error correction.
\newblock {\em IEEE Transaction on Information Theory}, 50(10):2315--2330,
  2004.

\bibitem{gottesman2014:LDPC}
D. Gottesman.
\newblock Fault-tolerant quantum computation with constant overhead.
\newblock {\em Quantum Information \& Computation}, 14(15-16):1338--1372, 2014.

\bibitem{PC08}
D.~Poulin and Y.~Chung.
\newblock On the iterative decoding of sparse quantum codes.
\newblock {\em Quantum Information \& Computation}, 8(10):987--1000, 2008.

\bibitem{delfosse2014:correlations}
N. Delfosse and J-P Tillich.
\newblock A decoding algorithm for {CSS} codes using the {X/Z} correlations.
\newblock In {\em Proc. of IEEE International Symposium on Information Theory,
  ISIT 2014}, pages 1071--1075, 2014.

\bibitem{leverrier2015:expander}
A. Leverrier, J-P Tillich, and G. Z{\'e}mor.
\newblock Quantum expander codes.
\newblock In {\em Foundations of Computer Science (FOCS), 2015 IEEE 56th Annual
  Symposium on}, pages 810--824. IEEE, 2015.

\end{thebibliography}

\newcommand{\SortNoop}[1]{}

\end{document}